\documentstyle[epsfig,floats,prb,aps]{revtex}
\begin{document}
\draft
\title{{\em Ab initio\/} zone-center phonons in LiTaO$_3$:
comparison to LiNbO$_3$}
\author{V.~Caciuc}
\address{
Universit\"at Osnabr\"uck -- Fachbereich Physik,
D-49069 Osnabr\"uck, Germany}
\author{A.~V.~Postnikov}
\address{
Gerhard Mercator University Duisburg -- Theoretical Low-Temperature Physics,
D-47048 Duisburg, Germany}
\date{May 11, 2001}
\twocolumn[\hsize\textwidth\columnwidth\hsize\csname@twocolumnfalse\endcsname
\maketitle
\begin{abstract}
The four $A_1$-TO $\Gamma$ phonon frequencies in lithium tantalate
are calculated in the frozen-phonon approach from first principles 
using the full-potential
linearized augmented plane wave method. A good agreement with the
experimental data available is found for all modes; 
reliable displacement pattern of different modes becomes
available from the calculated eigenvectors. The Raman spectra
recorded for $A_1$ modes in LiNbO$_3$  exhibit a counter-intuitive
softening of the $A_1$-TO$_3$ mode frequency with respect 
to that measured in
LiTaO$_3$. We explain this behavior by a comparatively harder oxygen
rotation in LiTaO$_3$ and discuss other differences in lattice dynamics
of two materials, notably delocalization of Ta and Li contributions
over more that one corresponding mode in LiTaO$_3$, differently
from the situation in lithium niobate. The Li isotope shift
is predicted in the calculation.
\end{abstract}
]
\section*{Introduction}

The ferroelectric materials LiNbO$_3$ and LiTaO$_3$ have been intensively
studied over years due to their promising applications in non-linear optical
and electro-optic devices -- see Ref.~\onlinecite{TAP61_131}
for a review on relevant properties of these materials. 
Both compounds possess rhombohedral space group $R3c$ 
($C_{3v}^6$) and have 10 atoms in the unit cell. 
They remain ferroelectric up to quite high transition temperatures:
$T_{\mbox{\tiny C}}$ amounts to 1480 K in LNbO$_3$
(that is probably the highest ferroelectric transition temperature
so far known) and 950 K in LiTaO$_3$.

Due to low symmetry and large number of atoms in the unit cell,
the first-principles studies of LiNbO$_3$ and LiTaO$_3$ proceed
not so fast as, e.g., for perovskite-type ferroelectric compounds,
KNbO$_3$ and KTaO$_3$ (the latter is an incipient ferroelectric, where
the transition can be induced by uniaxial pressure, or by doping).
Soon after first \emph{ab initio} studies of electronic 
structure\cite{PRB50_1992}, the analysis of
the aspects of ferroelectric instability
by Inbar and Cohen\cite{InbarCohen,PRB53_1193} was illuminating
in that they emphasized an importance of oxygen-rotating mode
in the ferroelectric transition. Earlier, the $z$-displacement of
Li ions out of the oxygen planes was believed to be the essence of 
the ferroelectric transformation.
Inbar and Cohen demonstrated that the energy profile in
the para- to ferroelectric transition (studied along the path
connecting experimentally determined end-point structures) is very
similar in both substances. As a possible explanation of the
difference in the $T_{\mbox{\tiny C}}$ values, they guessed
that the zone-boundary behavior may be found to be different 
in LiNbO$_3$ and LiTaO$_3$. This assumption has not yet been 
tested. 

While we do not access exactly this topic in the present study,
we believe that the analysis of phonon properties, even at
zone center, can give a clue to understanding different behavior
of LiNbO$_3$ and LiTaO$_3$.
Due to the complexity of structure, the analysis of lattice vibrations
in experiment was in part contradictory, and in what regards theory
only few calculations have been done. For LiNbO$_3$ we calculated
earlier\cite{JPCS61_295,PRB61_8806}
the frequencies and eigenvectors of TO-$\Gamma$ phonons,
based on a precision total-energy fit in the multidimensional
(up to dimension 9, for the $E$ block) space of symmetry coordinates.
Parlinski \emph{et al.} \cite{PRB61_272} calculated the phonon dispersion
over the full Brillouin zone of LiNbO$_3$, using the direct method and
the Fourier transformation of force constants sampled by 
displacing individual atoms in a 80-atoms (2$\times$2$\times$2)
supercell.
Semiempirical phonon calculations (i.e., fitting the
model parameters to experimentally measured frequencies)
have been done by Repelin \emph{et al.} \cite{JPCS60_819}
for both LiNbO$_3$ and LiTaO$_3$.
We are not aware of any first-principles phonon calculations 
for LiTaO$_3$.

%
%
\begin{table*}[t]
\caption{$A_1$-TO phonon frequencies
in LiNbO$_3$ and LiTaO$_3$}
\begin{tabular}{l*{10}{c}}
 & exp. LiNbO$_3$ && \multicolumn{4}{c}{exp. LiTaO$_3$} &&
LiTaO$_3$ & $^6$LiTaO$_3$ & (Li$\,^{m_{\mbox{\tiny Nb}}}$TaO$_3$) \\
\hline
Ref. & 
$\begin{array}{c}\mbox{[\protect\onlinecite{PRB56_5967} ]}\end{array}$ &&
$\begin{array}{c}\mbox{[\protect\onlinecite{PSSB142_287}]}\end{array}$ &
$\begin{array}{c}\mbox{[\protect\onlinecite{PRB38_10007}]}\end{array}$ &
$\begin{array}{c}\mbox{[\protect\onlinecite{JJAP32_4373}]}\end{array}$ & 
$\begin{array}{c}\mbox{[\protect\onlinecite{JPCS60_819} ]}\end{array}$ &&
\multicolumn{3}{c}{present calculation} \\
TO$_1$ & 252 && 206 & 203 & 206 & 201 && 194 & 199 & (197) \\
TO$_2$ & 275 && 253 & 252 & 253 & 253 && 242 & 252 & (261) \\
TO$_3$ & 332 && 356 & 356 & 356 & 356 && 360 & 361 & (360) \\
TO$_4$ & 632 && 600 & 597 & 597 & 597 && 599 & 599 & (602) \\
\end{tabular}
\label{tab:freq_A1}
\end{table*}

%
%
\begin{table*}[t]
\caption{Internal coordinates in LiTaO$_3$ (experiment and 
theoretical optimization)}
\begin{tabular}{l*{12}{c}}
&&
\multicolumn{3}{c}{Ta} &&
\multicolumn{3}{c}{Li} &&
\multicolumn{3}{c}{O} \\ 
\hline
Experiment: Ref.~\protect\onlinecite{JPCS34_521} &&
$\begin{array}{c}0\end{array}$&
$\begin{array}{c}0\end{array}$&
$\begin{array}{c}0\end{array}$&&
$\begin{array}{c}0\end{array}$&
$\begin{array}{c}0\end{array}$&
$\begin{array}{c}0.279\end{array}$&&
$\begin{array}{c}0.050\end{array}$&
$\begin{array}{c}0.344\end{array}$&
$\begin{array}{c}0.069\end{array}$\\
Present calculation  &&
$\begin{array}{c}0\end{array}$&
$\begin{array}{c}0\end{array}$&
$\begin{array}{c}0\end{array}$&&
$\begin{array}{c}0\end{array}$&
$\begin{array}{c}0\end{array}$&
$\begin{array}{c}0.282\end{array}$&&
$\begin{array}{c}0.049\end{array}$&
$\begin{array}{c}0.343\end{array}$&
$\begin{array}{c}0.071\end{array}$\\
\end{tabular}
\label{tab:IntCoord}
\end{table*}

On the experimental side, at least for the $A_1$ modes
there is very good agreement between different measurements
in what regards vibration frequencies -- see Table \ref{tab:freq_A1}.
Comparing them with the frequencies for LiNbO$_3$ shown
in the same table one can note that the frequencies are lower
in LiTaO$_3$ for all modes but one, the TO$_3$. Whereas the mode 
softening can be -- at least qualitatively -- intuitively related to
larger mass of Ta, and different degree of softening of
different modes -- to different participation of Ta (Nb) in them,
the hardening of the TO$_3$ mode on going from LiNbO$_3$
to LiTaO$_3$ is clearly counter-intuitive and can be only traced
to the differences in electronic properties.

The aim of the present study is to provide an \emph{ab initio}
description of zone-center $A_1$-TO phonons in LiTaO$_3$
and discuss them in comparison with corresponding phonons
in LiNbO$_3$, thus outlining small but important differences
in electronic properties in these materials. In the following
we briefly outline the details of calculation and proceed
with the discussion of phonon frequencies and eigenvectors.

\section*{Calculation scheme}

Our frozen-phonon calculation was based on precision
all-electron total-energy calculations with the use of
full-potential linearized augmented plane waves method,
as implemented in the WIEN97 code\cite{wien97}.
The local orbitals extension according to Ref.~\onlinecite{PRB43_6388}  
has been used for a better description of Ta $5s$ and $5p$,
Li $1s$ and O$2s$ states. 
The muffin-tin radii chosen were 1.88 a.u. for Ta and
1.65 a.u. for Li and O.
The exchange-correlation energy has been
treated in the local density approximation as parametrized
by Perdew and Wang\cite{PRB45_13244}. 

The Brillouin zone integrations were carried
out by the improved tetrahedron method\cite{PRB49_16223} on a
4$\times$4$\times$4 special $k$-points mesh which generated 13 $k$ points in
the irreducible Brillouin zone. When using a denser 6$\times$6$\times$6 k-mesh
(32 irreducible $k$-points), we found a difference in the total-energy trends
less than 1 mRy, negligible for the analysis of the lattice dynamics
in the scale of energy variations discussed below.
The planewave cutoff parameter
$R_{\mbox{\footnotesize min}}K_{\mbox{\footnotesize max}}$
for the basis was set to 9.5, that resulted in, on the average,
2313 basis functions for each $k$-point; the cutoff parameter
$G_{\mbox{\footnotesize max}}$ for the charge density expansion
was 12.0, i.e., these cutoff values had to be set to higher values
than in our earlier calculations for LiNbO$_3$\cite{PRB61_8806}
in order to achieve acceptable convergence.
The optimized lattice parameters turn out to be $a$=5.1274 a.u.
and $c$=13.7033 a.u., giving the error of $-2\%$ in the volume
and $-0.1\%$ in the $c/a$ ratio as compared to experimental values
$a$=5.1543, $c$=13.7835 a.u.\cite{JPCS34_521}. 
The subsequent calculations analyzing
the atomic displacements have been performed with the lattice parameters
fixed at their experimental values.

As regards the frozen-phonon formalism as such, it is known since
long\cite{KuncMar} and has been successfully applied to many systems.
This approach is physically transparent, and its accuracy is
easy to control. The bottleneck of the method is the necessity
to calculate the forces (or, alternatively, expansion coefficients
of the total energy over individual atomic displacements)
for many trial geometries, in order to map the total energy 
near equilibrium to quadratic or, if necessary, higher orders
in displacements.
This rapidly makes the computation unpractical for moderately
large systems, and/or those with reduced symmetry (where many
vibrational degrees of freedom mix in the same irreducible
representation). Earlier, we succeeded in calculating zone-center
phonons for, at the largest, the 9-dimensional $E$-block
of LiNbO$_3$\cite{PRB61_8806}, adding progressively data points
(i.e., trial geometries) that would refine normal coordinates,
as the latter emerge more and more clearly in the course
of calculations. In the present study, our concern is
the 4-dimensional $A_1$ block of the crystal vibrations in LiTaO$_3$,
for which we make a comparison with previously obtained 
LiNbO$_3$ data\cite{PRB61_8806}.
In order to obtain a reliable mapping of the total energy
we included 170 different geometries in our calculation.
The optimized internal coordinates as they emerge from the
four-dimensional total-energy fit are shown in Table \ref{tab:IntCoord}.
They serve as reference point for all displacements discussed later on.

\section*{Results}

For the set of four symmetry coordinates $S_{1{\cdots}4}$
chosen in terms of individual Cartesian displacements
as specified in Ref.~\onlinecite{PRB61_8806} for LiNbO$_3$,
the diagonalization of the dynamical matrix yields four
normal coordinates as follows:
\begin{eqnarray*}
\left(
\begin{array}{c}
S_{\mbox{\small TO$_1$}} \\
S_{\mbox{\small TO$_2$}} \\
S_{\mbox{\small TO$_3$}} \\
S_{\mbox{\small TO$_4$}} 
\end{array}
\right)
\! =
\! \left(
\begin{tabular}{r@{.}lr@{.}lr@{.}lr@{.}l}
\setlength{\tabcolsep}{1pt}
 -2&145 & -1&189 & -0&457 & ~1&120 \\
  3&915 & -1&371 &  0&131 &  0&265 \\
 -0&253 &  0&356 & -0&350 &  1&563 \\
 -1&039 & -0&157 &  1&153 &  0&373 
\end{tabular}
\right)
\! \left(
\begin{array}{c}
S_1 \\ S_2 \\ S_3 \\ S_4 
\end{array}
\right) \;.
\end{eqnarray*}

%
%
\begin{figure*}[th]
\centerline{\epsfig{file=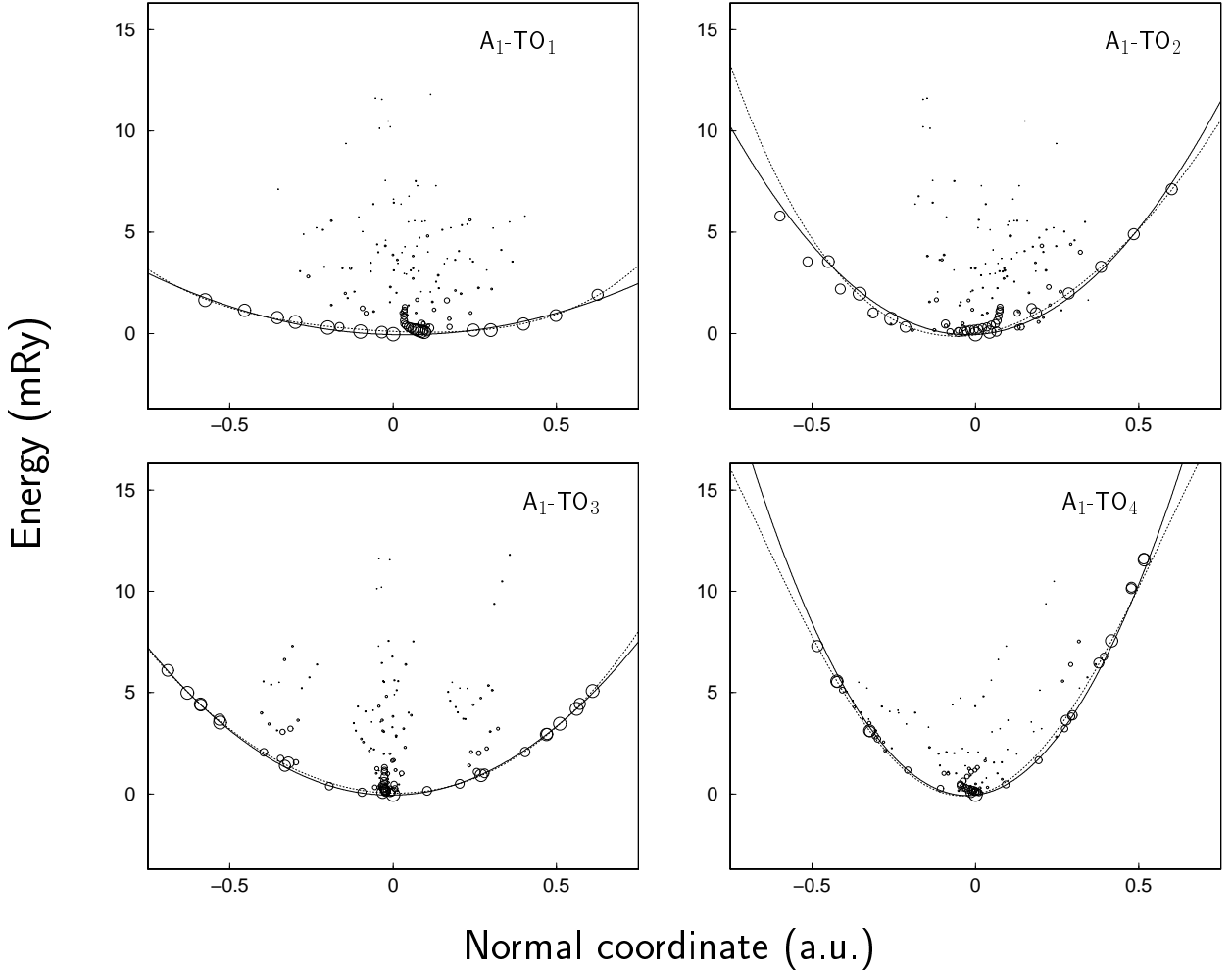,width=18.0cm}}
\caption{
The cut of the total-energy fit along the directions of normal displacements
in the four-dimensional space of symmetry coordinates corresponding to the
$A_1$ symmetry. The circles' sizes are inversely proportional to the distances
of corresponding data points from the direction of normal displacement. Solid
lines represent the 2d-order polynomial fit, dashed lines -- the 4th order fit.
}
\label{fig:EnergyCut}
\end{figure*}

These normal coordinates serve as abscissa in Fig.~\ref{fig:EnergyCut}
that visualizes corresponding cuts of the total energy hypersurface.
The number of trial configurations allows to perform a stable
4th order polynomial fit, also shown in Fig.~\ref{fig:EnergyCut}. 
It is obvious that the corrections to the 2d order fit are small
and not always physically sane, i.e. not all leading 
4th order terms are positive.
In other words, all four $A_1$-TO modes, according to our
calculation, are essentially harmonic, much more so than it was
the case for their counterparts in LiNbO$_3$ (see a similar
analysis in Ref.~\onlinecite{PRB61_8806}).
Another ``empirical'' justification of this fact is much better
agreement of measured and calculated (harmonic) frequencies than it was
the case in LiNbO$_3$ (see Table~\ref{tab:freq_A1}). The largest 
difference (7 -- 12 cm$^{-1}$) between our calculation and experiment
occurs for the TO$_1$ and TO$_2$ modes, with experimental values
higher, as it could be usually expected due to anharmonicity. For two other
modes, the agreement between experiment and theory is almost perfect,
as consistent with neat parabolic fit through data points along
the directions of these two normal coordinates in Fig.~\ref{fig:EnergyCut}.
In LiNbO$_3$, the difference between theory and experiment amounted
to 46 cm$^{-1}$ (TO$_1$) and 50 cm$^{-1}$ (TO$_4$ mode), both these modes
being considerably unharmonic judging on their corresponding 
total energy profiles.
In the following, we discuss the composition of different modes
and address the ``anomalous'' TO$_3$ mode (i.e., that is harder
in tantalate than in niobate) in some detail.

Since the LiNbO$_3$ and LiTaO$_3$ are in many aspects so similar,
we would like, when discussing their phonon frequencies,
to know whether the existing differences are mostly due to higher 
tantalum mass, or to variations in corresponding force constants. The simplest
illustration of the effect of mass is the calculation of frequencies
with the mass of Nb atom taken instead of that of Ta;
the results are given in the last column of Table~\ref{tab:freq_A1}.
Unexpectedly enough, the frequency noticeably increases 
in the course of such substitution solely for the TO$_2$ mode; 
for three other modes, even for the TO$_1$ (the mainly ``Nb'' mode
in LiNbO$_3$, see Ref.~\onlinecite{PRB61_8806}) the effect is
negligible, consistently with much smaller Ta contribution 
in corresponding eigenvectors (see below).

%
%
\begin{figure}[th]
\centerline{\epsfig{file=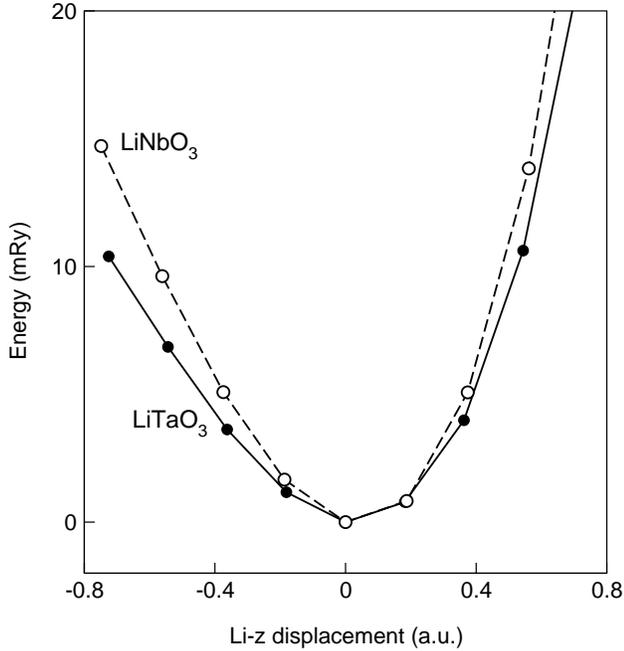,width=9.0cm,angle=-90}}
\bigskip
\caption{
Potential energy profile associated with Li z-displacements in $A_1$-TO$_3$
mode of LiNbO$_3$ (LNO) and LiTaO$_3$ (LTO). Positive movements shorten the
distance between Li and O ions, while the negative movements shorten that
between Li and Nb(Ta) ions.
}
\label{fig:LTO_LNO_Li}
\end{figure}

%
%
\begin{table}[b]
\caption{Distances (a.u.) to nearest neighbors and next nearest neighbors
of transition metal and oxygen to lithium atom in LiNbO$_3$
and LiTaO$_3$}
\begin{tabular}{lcr@{.}lr@{.}lcr@{.}lr@{.}l}
 & & \multicolumn{2}{c}{Nb(Ta)$_{\mbox{\tiny NN}}$}  &
     \multicolumn{2}{c}{Nb(Ta)$_{\mbox{\tiny NNN}}$} &
   & \multicolumn{2}{c}{ O$_{\mbox{\tiny NN}}$}  &
     \multicolumn{2}{c}{ O$_{\mbox{\tiny NNN}}$} \\
\hline
LiNbO$_3$ && 5&595 & 7&159 && 3&799 & 4&276 \\
LiTaO$_3$ && 5&669 & 7&355 && 3&848 & 4&367 \\
\end{tabular}
\label{tab:dist_Li}
\end{table}

Such ``Ta isotope'' effect roughly accounts for the half of 
the TO$_2$ frequency shift in Li(Ta$\rightarrow$Nb)O$_3$; 
the rest can only come from the steeper potential well associated
with Li $z$-displacement in LiNbO$_3$.
The corresponding total energy profile
is shown in Fig.~\ref{fig:LTO_LNO_Li}. The reason for the
``softness'' of the LiTaO$_3$ lattice with respect to such
Li-only vibration can be simply a somehow larger interatomic spacing,
as illustrated by Table \ref{tab:dist_Li}. It was already
emphasized earlier by Inbar and Cohen \cite{InbarCohen} 
that Li always remains quite well ionized in both tantalate
and niobate and that no effects of dynamical covalency 
that would lead to coupling of oxygen and lithium motions are seen 
in these materials. 

%
%
\begin{table*}[t]
\caption{Eigenvectors of TO phonons in LiNbO$_3$ and LiTaO$_3$ 
in the hexagonal setting}
\begin{tabular}{lr@{.}lr@{.}lc@{(}*{3}{r@{.}l}@{~)}
 @{\hspace*{7mm}}r@{.}lr@{.}lc@{(}*{3}{r@{.}l}@{~)}}
 & \multicolumn{11}{c}{LiTaO$_3$} & \multicolumn{11}{c}{LiNbO$_3$} \\
 & \multicolumn{2}{c}{~Ta}&\multicolumn{2}{c}{~Li}&\multicolumn{7}{c}{O} 
 & \multicolumn{2}{c}{~Nb}&\multicolumn{2}{c}{~Li}&\multicolumn{7}{c}{O}\\
\hline
TO$_1$ &    0&060 &    0&168 &&    0&096 &    0&132 & $-$0&104 &
            0&143 &    0&035 &&    0&036 &    0&085 & $-$0&123 \\ 
TO$_2$ &    0&109 & $-$0&191 && $-$0&047 &    0&038 & $-$0&080 &
            0&068 & $-$0&254 &&    0&015 & $-$0&014 &    0&001 \\ 
TO$_3$ & $-$0&007 &    0&051 && $-$0&441 & $-$0&101 & $-$0&003 &
            0&007 & $-$0&001 && $-$0&463 & $-$0&156 & $-$0&006 \\ 
TO$_4$ &    0&029 &    0&022 && $-$0&127 & $-$0&440 & $-$0&037 &
            0&023 &    0&015 && $-$0&082 & $-$0&437 & $-$0&022 \\ 
\end{tabular}
\label{tab:evec_A1}
\end{table*}
%
%
\begin{table*}[t]
\caption{Cartesian coordinates at equilibrium (top line)
and displacement patterns in the four $A_1$ modes
of LiNbO$_3$ and LiTaO$_3$, relative to the Ta(Nb) atom and
with the oxygen displacements normalized to 1.}
\begin{tabular}{lr@{.}l*{3}{cr@{.}lr@{.}lr@{.}l}}
 & \multicolumn{2}{c}{Li}  && \multicolumn{6}{c}{O1} &&
   \multicolumn{6}{c}{O2}  && \multicolumn{6}{c}{O3} \\
 & \multicolumn{2}{c}{$z$} &&
   \multicolumn{2}{c}{$x$} &
   \multicolumn{2}{c}{$y$} &  \multicolumn{2}{c}{$z$} &&
   \multicolumn{2}{c}{$x$} &
   \multicolumn{2}{c}{$y$} &  \multicolumn{2}{c}{$z$} &&
   \multicolumn{2}{c}{$x$} &
   \multicolumn{2}{c}{$y$} &  \multicolumn{2}{c}{$z$} \\
\hline
   \multicolumn{24}{c}{LiTaO$_3$} \\
coord. & 0&282$c$ && $-$0&280$a$ &    0&122$a$ & 0&405$c$ &&
                        0&034$a$ & $-$0&304$a$ & 0&405$c$ &&
                        0&246$a$ &    0&181$a$ & 0&405$c$ \\
TO$_1$ &    0&75 &&    0&13 & $-$0&04 & $-$0&38 &&
                    $-$0&04 &    0&13 & $-$0&38 &&
                    $-$0&10 & $-$0&10 & $-$0&38 \\
TO$_2$ & $-$1&14 &&    0&04 &    0&09 & $-$0&40 &&
                    $-$0&10 & $-$0&01 & $-$0&40 &&
                       0&05 & $-$0&08 & $-$0&40 \\
TO$_3$ &    0&22 && $-$0&09 &    0&40 & $-$0&00 &&
                    $-$0&30 & $-$0&28 & $-$0&00 &&
                       0&39 & $-$0&12 & $-$0&00 \\
TO$_4$ &    0&07 && $-$0&38 & $-$0&09 & $-$0&12 &&
                       0&27 & $-$0&28 & $-$0&12 &&
                       0&11 &    0&37 & $-$0&12 \\
\hline
   \multicolumn{24}{c}{LiNbO$_3$} \\
coord. & 0&281$c$ && $-$0&278$a$ &    0&127$a$ & 0&401$c$ &&
                        0&029$a$ & $-$0&304$a$ & 0&401$c$ &&
                        0&249$a$ &    0&177$a$ & 0&401$c$ \\
TO$_1$ & $-$0&01 &&    0&06 &    0&01 & $-$0&40 &&
                    $-$0&04 &    0&05 & $-$0&40 &&
                    $-$0&03 & $-$0&06 & $-$0&40 \\
TO$_2$ & $-$6&17 && $-$0&07 & $-$0&05 & $-$0&40 &&
                       0&07 & $-$0&03 & $-$0&40 &&
                    $-$0&01 &    0&08 & $-$0&40 \\
TO$_3$ & $-$0&01 && $-$0&14 &    0&38 & $-$0&02 &&
                    $-$0&27 & $-$0&31 & $-$0&02 &&
                       0&40 & $-$0&08 & $-$0&02 \\
TO$_4$ &    0&04 && $-$0&38 & $-$0&14 & $-$0&08 &&
                       0&31 & $-$0&26 & $-$0&08 &&
                       0&07 &    0&39 & $-$0&08 \\
\end{tabular}
\label{tab:disp_A1}
\end{table*}

Let us now turn to calculated eigenvectors shown in Table~\ref{tab:evec_A1}
in the hexagonal setting. The ``prototype'' oxygen atom is that
whose coordinates are given in Table~\ref{tab:IntCoord}; from its
eigenvector components $(x,y,z)$ the components related to other 
oxygen atoms can be reconstructed as 
$(\bar{y},x\!-\!y,z)$, $(y\!-\!x,\bar{x},z)$,
$(y\!-\!x,y,\frac{1}{2}+z)$, $(\bar{y},\bar{x},\frac{1}{2}+z)$, $(x,x\!-\!y,\frac{1}{2}+z)$.
For comparison, the eigenvector components for LiNbO$_3$ are also given;
their transformation to the Cartesian system according to
\begin{eqnarray*}
\left( \begin{array}{c} x\\y\\z \end{array} \right)
=\left( \begin{array}{ccc}
~0 & \sqrt{3}/2 & 0       \\
~1 & -1/2       & 0       \\
~0 & ~0         & c/a
\end{array} \right)
\left( \begin{array}{c} x_h\\y_h\\z_h \end{array} \right) \;.
\end{eqnarray*}
recovers the eigenvectors for LiNbO$_3$ as given in in
Ref.~\onlinecite{PRB61_8806}.
As in LiNbO$_3$, the heavy transition-metal 
ion contributes mostly to the two low-frequency modes. In the softest one,
it vibrates along with Li against the oxygen octahedra which are only
slightly tilted in this process (the tilting originates merely from
a crystal structure frustration, as was discussed 
in Ref.~\onlinecite{InbarCohen}). This is the ferroelectric mode,
frozen down from the paraelectric phase, near its ``freezing point''.
The TO$_2$ mode includes lithium vibration against both transition-metal
and oxygen sublattices; two other modes are too high for efficiently
resonating with transition-metal vibrations. 

The Li contribution exhibits some differences in two systems. 
To make these differences, somehow obscured in
center-of-mass related eigenvectors, more clear, we show
in Table~\ref{tab:disp_A1} pure (not scaled with masses)
vibration patterns of the four modes, relative to the heaviest 
Nb(Ta) atom and normalized so that the displacement of (any)
second heaviest constituent atom, i.e. oxygen, is 1. 
Whereas in LiNbO$_3$ the lithium vibration is essentially confined 
to the TO$_2$ mode (that alone is affected
by the Li isotope shift \cite{JJAP32_4373,JPCS61_295}), 
in LiTaO$_3$ the Li effect on the TO$_1$ and TO$_3$ modes
is also not negligible, with possible implications for
their frequencies on Li isotope doping.
The frequencies calculated with reduced Li mass are shown in
Table~\ref{tab:freq_A1}. The maximal frequency shift (for the TO$_2$ mode) 
reduces from 19 cm$^{-1}$ in LiNbO$_3$ to 10 cm$^{-1}$ in LiTaO$_3$. 
We guess that the Li isotope shift of the TO$_1$ mode is probably 
still large enough to be detected experimentally. 
Moreover, the Li-related phonon density of states,
if ever calculated similarly to how it was done 
in Ref.~\onlinecite{PRB61_272} for LiNbO$_3$, must certainly become
spread off lower and higher frequencies.

The second largest contribution of Li occurs, for both materials,
in the eigenvector of the TO$_1$ mode. Even if it is much smaller
(relative to the center of masses) in niobate, 
the displacement pattern of this mode is oxygen 
$z$-vibration against both the heaviest atom \emph{and} lithium,
so that the Li--O stretching is anyway present.
Here again the hardening of
the Li-$z$-related potential well on changing from LiTaO$_3$
to LiNbO$_3$ plays a role: the mere change of mass produces
only a minor effect in elevating the calculated (harmonic)
frequency to 208 cm$^{-1}$ in niobate\cite{PRB61_8806};
the strong anharmonicity of this mode contributes the rest
up to its measured frequency of 252 cm$^{-1}$.

The two hardest modes exhibit but negligible Nb(Ta) contribution
in both materials; the effect of Li $z$-displacement is also
quite reduced (although remains noticeable in LiTaO$_3$), so we have
here two almost exclusively O$_{xy}$ patterns, almost identical
in tantalate and niobate. These two modes are, correspondingly, 
a nearly rigid rotation of oxygen octahedra hosting Nb atoms 
and the bending of the octahedra (rotation of three
top atoms against the three bottom ones). The ``top view'' 
of corresponding distortions has been shown in Fig.~2 of 
Ref.~\onlinecite{JPCS61_295} for LiNbO$_3$.
What is noteworthy about TO$_3$ and TO$_4$ modes is 
that for \emph{both} of them
the calculated frequency is by $\sim$16 cm$^{-1}$ higher in tantalate.
Then, the anharmonicity of TO$_4$ in LiNbO$_3$ (but not in LiTaO$_3$)
accounts for a considerable increase of experimental frequency
in the niobate. The TO$_3$ mode is well harmonic
in both materials, so that the relation between phonon frequencies
in niobate and tantalate as predicted by the calculation 
is essentially retained in reality.
Since the TO$_4$ mode in both materials is almost free 
of transition-atom and lithium contributions,
the last panel of Fig.~\ref{fig:EnergyCut}
and the corresponding energy profile in Fig.~1 
of Ref.~\onlinecite{PRB61_8806} exhibit essentially 
the effect of bending of oxygen octahedra
and give a fair indication that the latter
are more rigid in LiTaO$_3$. 

An analogy to that can be found in cubic perovskites: among
many calculations done by date of zone-center phonons in
KNbO$_3$ and KTaO$_3$ by comparable 
techniques\cite{SB92,PRB53_176,PRB50_758,PRL74_4067}, 
the calculated frequencies of two
hardest modes, that correspond to different distortions 
of oxygen octahedra, are higher in tantalate than in niobate.
The experimental situation confirms this at least for the
TO$_4$ mode (the TO$_3$ of the $T_{2u}$ symmetry is Raman 
and infrared silent in cubic perovskite, so that its measurements 
are not numerous; corresponding frequencies any anyway quite close 
in KNbO$_3$ and KTaO$_3$).
The possible origin of this seemingly systematic trend
can be sought for either in a slightly increased lattice spacing
of tantalates -- that would however probably have an opposite
effect on the oxygen rigidity, if at all -- or in a slight
difference in the degree of covalency. The latter guess gets
support from the analysis of the dynamical matrix corresponding to
the soft-mode motion (i.e., also including the oxygen tilting)
by Inbar and Cohen\cite{PRB53_1193}. They have shown that
the ``full'' O--O elements of this matrix are larger in LiTaO$_3$, 
whereas the relation between Madelung contributions to the dynamical
matrix (i.e., those assuming nominal ionic charges) is opposite. 

The aspects of relative covalency of tantalates and niobates
are influenced by several factors; discussions to the point 
can be found in Ref.~\onlinecite{PRB53_176} on perovskites and in
Ref.~\onlinecite{PRB53_1193} on our present systems of interest.
Summarizing, the covalency seems to be higher in tantalates, judging by
a slightly wider valence band and a stronger dispersion of certain
bands, that is seemingly related to less localized $d$-function of tantalum.
A possible link to the stiffness of oxygen octahedra may be
that higher covalency enhances interaction between lattice-dynamical
degrees of freedom, softening the softest modes
and hardening the hardest. Indeed, the span of TO-frequencies,
at least in the harmonic approximation, seems to be typically
broader in tantalates. Different degree of anharmonicity
of different modes may complicate the experimental situation.

\section*{Conclusion}

We extended our previous frozen-phonon analysis of zone center-TO
phonons in lithium niobate over lithium tantalate. For all $A_1$ modes
quite good agreement was found between the
experimental and calculated phonon frequencies, indicating, differently
from LiNbO$_3$, a quite harmonic nature of all four modes.
In spite of large similarity in crystal structure and electronic
properties, the phonon eigenvectors show certain differences in
both systems. Specifically, the Li contribution, that was
in LiNbO$_3$ only noticeable in the single TO$_2$ mode, now
spreads over TO$_1$ to TO$_3$ modes. One can expect the Li isotope
shift in LiTaO$_3$ to become less pronounced but affecting 
these three modes. Li displacement experiences a softer
potential well in LiTaO$_3$, due to more expanded lattice, 
that lowers the frequencies of Li-dependent modes.
The contribution of the transition metal ion
affects the TO$_2$ mode more than TO$_1$, contrary to the case
in LiNbO$_3$. 
Contrary to the effect on Li, the rotational modes of
oxygen octahedra are noticeably hardened in LiTaO$_3$,
indicating a more stiff oxygen network. Due to this, a higher
frequency of the TO$_3$ mode in tantalate is explained.
For the TO$_4$ mode, the harmonic part of which also gets harder
in LiTaO$_3$, the experimental situation is reversed only
due to a strong anharmonicity of this mode in LiNbO$_3$.

\section*{Acknowledgments}
The work was supported by the German Research Society (graduate
college ''Microstructure of oxidic crystals''). 
The authors are grateful to G. Borstel for introduction to the subject. 
A.V.P. appreciates stimulating discussions with
P.~Bourson, M.~Fontana and V.~Vikhnin.

\end{document}